\documentclass[prl,aps,amsfonts,amssymb,showpacs,nofootinbib,floats,twocolumn]{revtex4}
\usepackage{graphics}
\usepackage{epsfig}
\usepackage{bm}

\begin{document}
\title{First-Order Type Effects in ${\rm
YBa_{2}Cu_{3}O_{6+x}}$ at the Onset of Superconductivity}
\author{L. Tassini}
\author{W. Prestel}
\author{A. Erb}
\author{M. Lambacher}
\author{R. Hackl}
\affiliation{Walther Meissner Institut, \linebreak Bayerische
Akademie der Wissenschaften, 85748 Garching, Germany}
\date{080701}

\begin{abstract}
We present results of Raman scattering experiments on tetragonal
${\rm (Y_{1-y}Ca_{y})Ba_{2}Cu_{3}O_{6+x}}$ for doping levels
$p(x,y)$ between 0 and 0.07 holes/CuO$_2$. Below the onset of
superconductivity at $p_{\rm sc1} \approx 0.06$, we find evidence
of a diagonal superstructure. At $p_{\rm sc1}$, lattice and
electron dynamics change discontinuously with the charge and spin
properties being renormalized at all energy scales. The results
indicate that charge ordering is intimately related to the
transition at $p_{\rm sc1}$ and that the maximal transition
temperature to superconductivity at optimal doping $T_{c}^{\rm
max}$ depends on the type of ordering at $p>p_{\rm sc1}$.

\end{abstract}

\maketitle

In cuprates the maximal transition temperature to
superconductivity $T_{c}^{\rm max}$ depends on the compound class.
In contrast, the variation of $T_{c}$ with doping $p$ does not,
and superconductivity exists  between  approximately 0.05  and
0.27  holes per $\rm CuO_2$ formula unit in clean samples
\cite{Tallon:1995}. In the presence of disorder this range shrinks
\cite{Tallon:1995} leading to a sample-specific onset point of
superconductivity at $p_{\rm sc1} \ge 0.05$. In addition to
superconductivity, short-range antiferromagnetism with the domains
separated by quasi one-dimensional charged stripes can occur
\cite{Cheong:1991,Tranquada:1995,Bianconi:1996,Kivelson:2003,Tranquada:2005}.
In ${\rm La_{2-x}Sr_xCuO_4}$ ($p=x$), this superstructure is
oriented along the diagonals of the CuO$_2$ plane below $p_{\rm
sc1}$ 
and rotates by $45^{\circ}$ at $p_{\rm sc1}$ \cite{Wakimoto:1999}.
This rotation was also seen in the low-energy electronic Raman
spectra where the ordering-related response flips symmetry
\cite{Tassini:2005}.

For $p > p_{\rm sc1}$ superstructures are observed in all cuprates
\cite{Tranquada:1995,Hoffman:2002,Hinkov:2004,Tranquada:2005,Stock:2005,Hinkov:2007}.
However, the type of ordering and its relationship to
superconductivity is rather complicated to pin down
\cite{Castellani:1998,Klauss:2000,Kivelson:2003}. In a few
compounds the lattice stabilizes static spin and charge
superstructures and the superconducting transition temperature is
reduced or quenched \cite{Tranquada:1995,Tranquada:2005}. In most
of the cases fluctuating order prevails, and it is particularly
hard to detect the charge part
\cite{Castellani:1998,Kivelson:2003}. Raman spectroscopy was found
to be a viable method \cite{Tassini:2005}.

\begin{figure}[b]
  \centering
  \includegraphics [width=6.5cm]{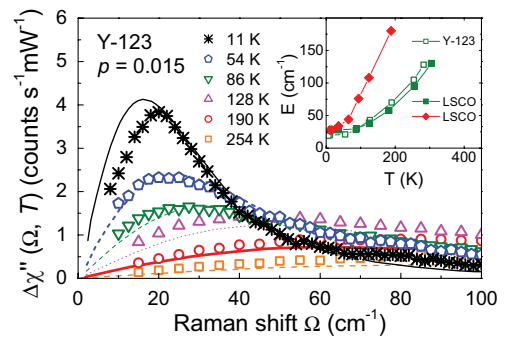}
  \caption[]{(Color online) Response of charge-ordering fluctuations in ${\rm
  (Y_{0.97}Ca_{0.03})Ba_{2}Cu_{3}O_{6.05}}$ \cite{Caprara:2005}.
  The points represent the experimental results, the full lines
  are theoretical fits. (Reproduced with permission.) Two fluctuations with finite but opposite
  momenta are exchanged (Azlamazov-Larkin diagrams). The inset
  shows the peak positions of the response as a function of temperature in
  ${\rm (Y_{0.97}Ca_{0.03})Ba_{2}Cu_{3}O_{6.05}}$, ${\rm La_{1.98}Sr_{0.02}CuO_4}$, and
  ${\rm La_{1.90}Sr_{0.10}CuO_4}$.  At similar doping ($p \simeq 0.02$) the points essentially coincide.

  }
\label{AL}
\end{figure}

Inelastic (Raman) scattering of light is capable of probing most
of the excitations in a solid including lattice vibrations, spins,
and electrons, as well as their interactions
\cite{Devereaux:2007}. Since the polarizations of the incident and
the scattered photons can be adjusted independently, many of the
excitations can be sorted out via the selection rules. For
instance, the transport properties of conduction electrons can be
measured independently in different regions of the Brillouin zone
\cite{Devereaux:2007,Opel:2000}, and the orientation of
(fluctuating) charged stripes can be determined
\cite{Tassini:2005}. A detailed model calculation
\cite{Caprara:2005} demonstrated that in addition to the symmetry
selection rules the dependence on energy and temperature of the
response related to stripes can be understood quantitatively
(Fig.~\ref{AL}) in terms of charge-ordering fluctuations.

In this paper, we focus on the ``high-$T_c$'' compound ${\rm
YBa_{2}Cu_{3}O_{6+x}}$ (Y-123) at doping levels $0 \le p \le
0.07$. The purpose is to gain insight into the nature of the onset
of superconductivity at $p_{\rm sc1} \approx 0.06$ and into
possible discrimination criteria to the ``low-$T_c$'' compound
${\rm La_{2-x}Sr_xCuO_4}$ (LSCO) studied earlier
\cite{Wakimoto:1999,Tassini:2005,Caprara:2005}. For $0<p<p_{\rm
sc1}$ we find spectral features strikingly similar to those in
LSCO and conclude that the diagonal
stripe-like superstructure is universal. Above $p_{\rm sc1}$, the
indications of the diagonal superstructure disappear. However, in
clear contrast to LSCO no indications of stripe ordering are found
in Y-123. Hence, the first-order type changes at $p_{\rm sc1}$ are
not a peculiarity of LSCO \cite{Wakimoto:1999,Tassini:2005} but
exist also, though with distinct differences, in Y-123. Along with
that, the direct spin-spin exchange interaction and the
electron-phonon coupling discontinuously de- and increase,
respectively, across $p_{\rm sc1}$.

\begin{figure*}
  \centering
  \includegraphics [width=12cm]{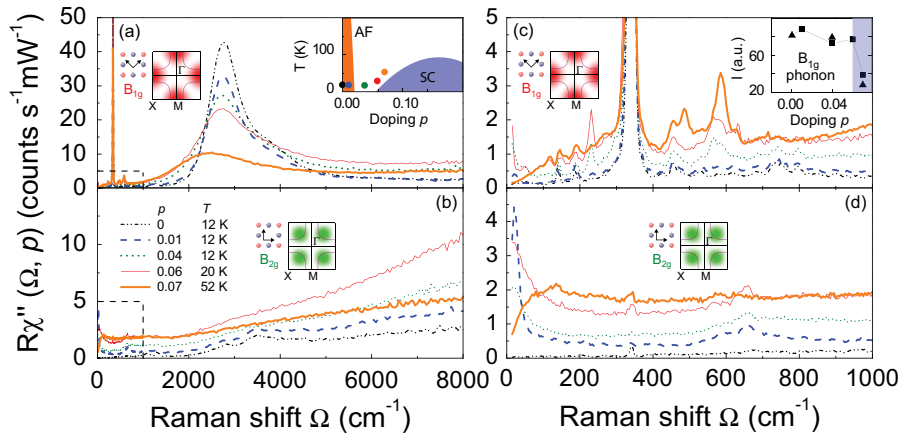}
  \vspace{-0.2cm}
  \caption[]{(Color online) Raman response $R\chi^{\prime
  \prime}(p,\Omega)$ of ${\rm (Y_{1-y}Ca_{y})Ba_{2}Cu_{3}O_{6+x}}$
  at low doping and temperature outside the superconducting state (raw data).
  The response $R\chi^{\prime\prime}(\Omega,p)$ is obtained by dividing the experimental
  spectra by the Bose factor.
  All spectra are measured with the Ar laser line at 458 nm.
  The upper (a,c) and lower (b,d) panels
  correspond to the $B_{1g}$ and $B_{2g}$ symmetries, respectively, with the
  related polarizations and sensitivities in the Brillouin zone as indicated.
  Doping level and measuring temperatures of the samples are shown as full circles
  in the schematic phase diagram in the inset in (a),
  where AF and SC label antiferromagnetism and superconductivity,
  respectively, and are indicated explicitly in (b).
  Panels (a) and (b) show an energy range of
  8000~cm$^{-1}$ equivalent to 1~eV. In (c) and (d) the low-energy
  and intensity range is expanded as indicated by dashed boxes in (a) and (b).
  The inset in (c) shows the peak intensity of the
  $B_{1g}$ ``buckling'' phonon at 340~cm$^{-1}$ (43~meV). Triangles and squares
  represent Ca-free and Ca-doped Y-123, respectively.
  }
\label{long-short-0}
\end{figure*}

For the studies we used single crystals of ${\rm
(Y_{1-y}Ca_{y})Ba_{2}Cu_{3}O_{6+x}}$ grown in BaZrO$_3$
cru\-cibles \cite{Erb:1996}. Doping with Ca allows us to directly
control the number of holes. For $x\simeq 0$, doping levels up to
$p=0.06$ are given by the Ca content, $p=y/2$ \cite{Tallon:1995}.
Crystals with 0, 2, 8, and 12\% Ca were studied, all not
superconducting for $x\simeq 0$. Superconductivity could be
induced only by oxygen co-doping ($y=0.08$, $x=0.3$), and $p
\simeq 0.07$ is estimated from the transition temperature
$T_c=28$~K via the universal relationship $T_c(p)$
\cite{Tallon:1995}. The material remains tetragonal on the
average. Hence, all samples studied here have the same
crystallographic structure. We made sure that the observed effects
are independent of the way of doping by also studying Ca-free
crystals with $x=0.3$ and 0.4 with $T_c=0$ and 26~K, respectively
[see Ref.~\cite{Tassini:2008} and inset of
Fig.~\ref{long-short-0}~(c)]. For $x=0.4$, the superconducting
sample is orthorhombic with the twin boundaries visible under
polarized light. The results are identical to within the
experimental uncertainty and will be published elsewhere.

In Figure~\ref{long-short-0} we show normal-state Raman spectra of
Y-123 at low temperature in the doping range $0 \leq p \leq 0.07$.
We describe first the overall trends in the energy range of
8000~cm$^{-1}$ (1~eV).

In $B_{1g}$ symmetry [Fig.~\ref{long-short-0}~(a)] an intense
phonon at 340~cm$^{-1}$ and scattering from nearest-neighbor
spin-flip excitations in the range between 2000 and 4000~cm$^{-1}$
are the most prominent features observed. The evolution with
doping of the magnetic scattering shows features beyond those
found before \cite{Sugai:2003,Gozar:2005}. As long as the doping
is below the onset point of superconductivity ($p \le p_{\rm
sc1}$) the peak height changes slowly and continuously while the
position is essentially constant. Upon crossing $p_{\rm sc1}$ the
two-magnon peak moves discontinuously downwards by approximately
250~cm$^{-1}$. Independent of doping, positions and intensities of
the two-magnon peaks react only mildly when the temperature is
raised \cite{Tassini:2008}.

In $B_{2g}$ symmetry [Fig.~\ref{long-short-0}~(b)] light
scattering from spin excitations is weak \cite{Devereaux:2007}. A
low-energy continuum appears only at finite carrier
concentrations. The high-energy response increases continuously
until superconductivity sets in. Then, the spectrum assumes a more
convex shape and is depressed at high energy.

Zooming in on low-energies [Fig.~\ref{long-short-0}~(c) and~(d)]
we observe a continuous increase of the electronic response upon
doping in $B_{1g}$ symmetry [Fig.~\ref{long-short-0}~(c)] with no
significant changes across $p=p_{\rm sc1}$. On the other hand, the
peak height of the $B_{1g}$ phonon at 340~cm$^{-1}$ collapses by
factor of two for $p>p_{\rm sc1}$ [inset of
Fig.~\ref{long-short-0}~(c)].

In $B_{2g}$ symmetry [Fig.~\ref{long-short-0}~(d)] we find a
complete suppression of the electronic Raman spectra below
400~cm$^{-1}$ for $p=0$ as expected for an insulator. As soon as
carriers are added the response becomes finite and a new peak in
the range 15--30~cm$^{-1}$ pops up similar to those in ${\rm
La_{1.98}Sr_{0.02}CuO_4}$ \cite{Tassini:2005} and ${\rm
(Y_{0.97}Ca_{0.03})Ba_{2}Cu_{3}O_{6.05}}$ (Fig.~\ref{AL}, \cite{Caprara:2005}). Up
to $p_{\rm sc1}$ the integrated cross section increases
proportional to $p$. For $p=0.07$ it does not change any further.

\begin{figure}[t]
  \centering
  \vspace{-0.1cm}
  \includegraphics [width=6cm]{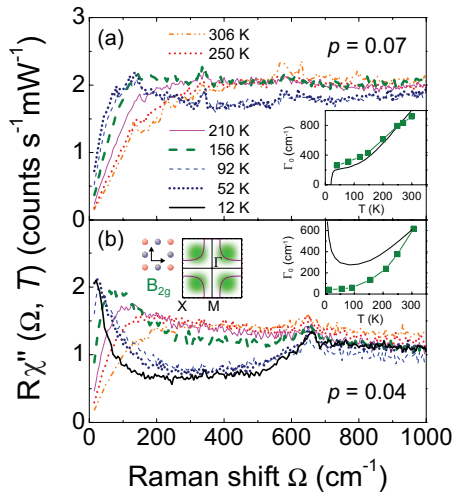}
  \vspace{-0.2cm}
  \caption[]{(Color online) Raman response $R\chi_{B_{2g}}^{\prime
  \prime}(T,\Omega)$ of  $\rm (Y_{1-y}Ca_{y})Ba_2Cu_3O_{6+x}$
  at $p=0.07$ (a) and 0.04 (b). In the insets we plot the
  low-energy ``Raman resistivities''
  $\Gamma_0(T)\propto(\partial\chi^{\prime\prime}/{\partial\Omega})^{-1}_{\Omega=0}$
  (squares) corresponding
  to the initial slope of the spectra \cite{Opel:2000} and compare
  them with the conventional resistivities (lines) from ref.~\cite{Ando:2004}.}
  \vspace{-0.4cm}
\label{short-T}
\end{figure}

In Fig.~\ref{short-T} we plot the low-energy $B_{2g}$ spectra at
two characteristic doping levels below and above $p_{\rm sc1}$ as
a function of temperature in order to further highlight the
qualitative changes across the critical doping. At $p=0.07$ the
response develops as expected from the resistivity and the results
at $p \ge 0.1$ \cite{Opel:2000,Venturini:2002,Ando:2004} including
a pseudogap of approximately 1000~cm$^{-1}$ opening up below 150~K
\cite{Opel:2000}. As shown in the inset of Fig.~\ref{short-T}~(a)
the Raman relaxation rate
$\Gamma_0(T)\propto(\partial\chi^{\prime\prime}/{\partial\Omega})^{-1}_{\Omega=0}$ at low energies follows
the resistivity when properly converted \cite{Opel:2000}. This is
in clear contrast to the non-superconducting sample at $p=0.04$
where a qualitative difference  between the Raman and the
transport resistivities is found in the entire temperature range
[inset of Fig.~\ref{short-T}~(b)]. As can be seen in the main
panel of Fig.~\ref{short-T}~(b) the discrepancy originates from
the low-energy peak developing below approximately 250~K. Along
with the peak a pronounced pseudogap of approximately
650~cm$^{-1}$ opens up which, if taken alone, could essentially
account for the increase of the conventional resistivity towards
low temperature.
The pivotal question is as to what drives the abrupt changes
versus doping level in the magnon, charge, and phonon spectra. In
what follows we show that these changes are related to the
superstructure in a universal way. So far, the rotation at
$p\simeq p_{\rm sc1}$ of the essentially 1D or stripe-like pattern
from diagonal to parallel with respect to the Cu--O bond direction
is only established in LSCO
\cite{Wakimoto:1999}, for which $T_c^{\rm max} \simeq 40$~K. In
the compounds with $T_c^{\rm max}$ in the 100~K range evidence of
ordering is found only at $p \ge 0.10$. In Y-123
\cite{Hinkov:2004,Stock:2005} and ${\rm
Bi_{2}Sr_{2}CaCu_{2}O_{8+\delta}}$ \cite{Hoffman:2002} the order
is more of checkerboard or 2D rather than of 1D or stripe type.
For $p_{\rm sc1}<p<0.1$, the superstructure in Y-123 seems to
disappear \cite{Stock:2006}. Very recently, nematic order was
proposed to occur close to $p \simeq 0.07$ \cite{Hinkov:2007}. To
our knowledge, there are no experiments below ${p_{\rm sc1}}$.

Here, we find collective modes in the Raman spectra at $0<p<p_{\rm
sc1}$ exhibiting shapes as well as symmetry and temperature
dependences similar to those in LSCO
\cite{Tassini:2005}. Apparently, they originate from the same
fluctuating spin and charge superstructure with a modulation along
the diagonals of the ${\rm CuO_2}$ plane
(Fig.~\ref{AL}) \cite{Caprara:2005}.
We are aware that our information is not on the structure directly
but only via the dynamics. However, there is no structural
analysis available and the selection rules are a particularly
strong argument as to symmetry breaking and orientation. In
addition, shape and temperature dependence demonstrate the
similarity of the underlying physics (Fig.~\ref{AL}). Hence, we
find indications of spin and charge ordering in a completely
tetragonal cuprate in the doping range between the antiferromagnet
and the onset of superconductivity. The type of ordering at
$0<p<p_{\rm sc1}$ is universal and depends neither on the material
class nor on structural details of the crystals such as lattice
distortions nor on the way of doping. In fact, Ca and O doping in
Y-123 lead to equivalent results \cite{Tassini:2008} [for the
$B_{1g}$ phonon see inset of fig.~\ref{long-short-0}~(c)]. In
particular, Ca is sitting on a site with the full lattice symmetry
and does not trap charges \cite{Janossy:2003}. Therefore, the
influence on the direction of the superstructure is expected to be
weaker than that of Sr in LSCO
\cite{Sushkov:2005} if not negligible. Alternatively, the shape of
the Fermi surface can influence the orientation of the ordering in
a fashion similar to nesting in materials with charge density
waves \cite{Gruner:1994}. With increasing doping the distance
between the stripes decreases \cite{Tranquada:2005, Wakimoto:1999}
and the ordering wave vector ${\bf q}_{\rm charge}$ becomes large
enough to connect the flat parts of the underlying Fermi surface
[see, e.g., inset of Fig.~\ref{long-short-0}~(c)]. Therefore,
diagonal order becomes less favorable for phase space reasons and
disappears above $p_{\rm sc1}$. Defects may shift $p_{\rm sc1}$
\cite{Tallon:1995}.

\begin{figure*}
  \centering
  \includegraphics [width=9cm]{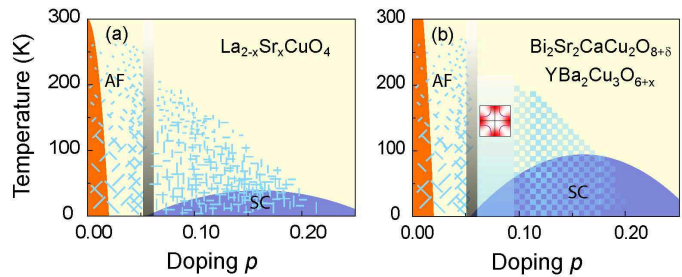}
  \vspace{-0.2cm}
  \caption[]{(Color online) Phase diagrams for ``low'' (a)
  and ``high''-$T_c$ (b) cuprates as
  derived from earlier \cite{Tassini:2005} and present Raman data and from
  neutron scattering
  \cite{Hinkov:2004,Tranquada:2005,Stock:2005,Stock:2006} and
  scanning tunneling microscopy
  \cite{Hoffman:2002,Kivelson:2003} measurements. The position of the onset point of
  superconductivity $p_{\rm sc1}$ depends on the degree of disorder
  with the minimum at approximately 0.05 for clean materials \cite{Tallon:1995}. The
  differently long lines qualitatively indicate doping and temperature ranges, density, orientation, and
  correlation lengths of stripes. The square pattern symbolizes
  checkerboard order which, however, is only established for $p>0.10$.
  For $p_{\rm sc1} < p < 0.10$, there are indications from the
  Raman results presented here for an
  interaction with structure in momentum space as sketched.}
\label{phase-diagram}
\end{figure*}

While symmetry arguments and analogies lead to straightforward
conclusions for $p < p_{\rm sc1}$ the analysis at $p > p_{\rm
sc1}$ is less direct. In contrast to LSCO,
where the ordering-induced collective modes change symmetry from
$B_{2g}$ to $B_{1g}$ along with the rotation of the superstructure
\cite{Tassini:2005}, we do not observe any ordering-related peaks
in Y-123 for $p>p_{\rm sc1}$ [see Fig.~\ref{long-short-0}~(c),
$p=0.07$ and Ref.~\cite{Tassini:2005}]. However, all the
renormalization effects described above demonstrate indirectly a
very strong interaction peaked close to the M points, i.e. with
$|d_{x^2-y^2}|$ symmetry, [see inset of
Fig.~\ref{long-short-0}~(c)] to become effective  at $p>p_{\rm
sc1}$. This interaction leads to a shift of the two-magnon peak
indicating a reduction of the Heisenberg exchange coupling $J$ by
almost 10\%. The related fluctuations also open a new coupling
channel of proper symmetry for the $B_{1g}$ phonon as indicated by
the discontinuous reduction of the peak intensity and the onset of
a (weak) Fano-type line shape. The effect on the ${\bf q} \simeq
0$ Raman phonon is relatively weak in contrast to the strong
coupling at ${\bf q}_{\rm charge} \simeq 0.25 (2\pi/a)$
\cite{Reznik:2006}. In fact, it would be interesting to study the
buckling and the half-breathing phonons close to the expected
${\bf q}_{\rm charge}$ right above and below $p_{\rm sc1}$.

In conclusion, superconductivity and the apparently universal
diagonal order at $p<p_{\rm sc1}$ are mutually exclusive.
On the
basis of the so far available results (including ours) it seems
that details of the order at $p>p_{\rm sc1}$ are crucial for
$T_{c}^{\rm max}$. In the ``low''-$T_c$ materials such as LSCO collinear stripe order prevails
\cite{Cheong:1991,Tranquada:1995,Bianconi:1996,Wakimoto:1999,Tranquada:2005}.
In compounds with $T_c$-s in the 100~K range the order is more 2D,
e.g. of checkerboard or nematic type, aligned with the $\rm CuO_2$
plane \cite{Hoffman:2002,Hinkov:2004,Stock:2005,Hinkov:2007}. For
$p_{\rm sc1} < p < 0.1$ we found a strongly momentum dependent
interaction which is symmetry compatible with the order at
$p>0.10$ and with the superconducting gap. Possible candidates are
quasicritical fluctuations of charge stripes \cite{Perali:2000} or
a fluctuating Fermi surface deformation
\cite{Metzner:2003,Hinkov:2007}. Either the electron-lattice
interaction \cite{Klauss:2000, Sushkov:2005} or details of the band structure
\cite{Seibold:2006} can influence the order. Combining our results
with earlier ones we arrive at the phase
diagrams sketched in Fig.~\ref{phase-diagram}.\\

{\bf Acknowledgements} Discussions with T.P. Devereaux, C.
Di~Castro, and M. Grilli are gratefully acknowledged. The project
has been supported by the DFG under grant numbers Ha2071/3 and
Er342/1 via the Research Unit FOR~538 .

%

\begin{thebibliography}{28}
\expandafter\ifx\csname natexlab\endcsname\relax\def\natexlab#1{#1}\fi
\expandafter\ifx\csname bibnamefont\endcsname\relax
  \def\bibnamefont#1{#1}\fi
\expandafter\ifx\csname bibfnamefont\endcsname\relax
  \def\bibfnamefont#1{#1}\fi
\expandafter\ifx\csname citenamefont\endcsname\relax
  \def\citenamefont#1{#1}\fi
\expandafter\ifx\csname url\endcsname\relax
  \def\url#1{\texttt{#1}}\fi
\expandafter\ifx\csname urlprefix\endcsname\relax\def\urlprefix{URL }\fi
\providecommand{\bibinfo}[2]{#2}
\providecommand{\eprint}[2][]{\url{#2}}

\bibitem[{\citenamefont{Tallon et~al.}(1995)\citenamefont{Tallon, Bernhard,
  Shaked, Hitterman, and Jorgensen}}]{Tallon:1995}
\bibinfo{author}{\bibfnamefont{J.~L.} \bibnamefont{Tallon}},
  \bibinfo{author}{\bibfnamefont{C.}~\bibnamefont{Bernhard}},
  \bibinfo{author}{\bibfnamefont{H.}~\bibnamefont{Shaked}},
  \bibinfo{author}{\bibfnamefont{R.~L.} \bibnamefont{Hitterman}},
  \bibnamefont{and} \bibinfo{author}{\bibfnamefont{J.~D.}
  \bibnamefont{Jorgensen}},
  \bibinfo{journal}{Phys. Rev. B}
  \textbf{\bibinfo{volume}{51}}, \bibinfo{pages}{12911} (\bibinfo{year}{1995}).



\bibitem[{\citenamefont{Cheong et~al.}(1991)}]{Cheong:1991}
\bibinfo{author}{\bibfnamefont{S.-W.} \bibnamefont{Cheong}},
  \bibinfo{author}{\bibfnamefont{G.}~\bibnamefont{Aeppli}},
  \bibinfo{author}{\bibfnamefont{T.~E.}~\bibnamefont{Mason}},
  \bibinfo{author}{\bibfnamefont{H.} \bibnamefont{Mook}},
  \bibinfo{author}{\bibfnamefont{S.~M.} \bibnamefont{Hayden}},
  \bibinfo{author}{\bibfnamefont{P.~C.} \bibnamefont{Canfield}},
  \bibinfo{author}{\bibfnamefont{Z.} \bibnamefont{Fisk}},
  \bibinfo{author}{\bibfnamefont{K.~N.} \bibnamefont{Clausen}},
  \bibnamefont{and} \bibinfo{author}{\bibfnamefont{J.~N.}
  \bibnamefont{Martinez}},
  \bibinfo{journal}{Phys. Rev. Lett.}
  \textbf{\bibinfo{volume}{67}}, \bibinfo{pages}{1791} (\bibinfo{year}{1991}).



\bibitem[{\citenamefont{Tranquada et~al.}(1995)\citenamefont{Tranquada,
  Sternlieb, Axe, Nakamura, and Uchida}}]{Tranquada:1995}
\bibinfo{author}{\bibfnamefont{J.~M.} \bibnamefont{Tranquada}},
  \bibinfo{author}{\bibfnamefont{B.~J.} \bibnamefont{Sternlieb}},
  \bibinfo{author}{\bibfnamefont{J.~D.} \bibnamefont{Axe}},
  \bibinfo{author}{\bibfnamefont{Y.}~\bibnamefont{Nakamura}}, \bibnamefont{and}
  \bibinfo{author}{\bibfnamefont{S.}~\bibnamefont{Uchida}},
  \bibinfo{journal}{Nature} \textbf{\bibinfo{volume}{375}},
  \bibinfo{pages}{561} (\bibinfo{year}{1995}).

\bibitem[{\citenamefont{Bianconi et~al.}(1996)}]{Bianconi:1996}
\bibinfo{author}{\bibfnamefont{A.} \bibnamefont{Bianconi}},
  \bibinfo{author}{\bibfnamefont{N.~L.}~\bibnamefont{Saini}},
  \bibinfo{author}{\bibfnamefont{A.}~\bibnamefont{Lanzara}},
  \bibinfo{author}{\bibfnamefont{M.} \bibnamefont{Missori}},
  \bibinfo{author}{\bibfnamefont{T.} \bibnamefont{Rossetti}},
  \bibinfo{author}{\bibfnamefont{H.} \bibnamefont{Oyanagi}},
  \bibinfo{author}{\bibfnamefont{H.} \bibnamefont{Yamaguchi}},
  \bibinfo{author}{\bibfnamefont{K.} \bibnamefont{Oka}},
  \bibnamefont{and} \bibinfo{author}{\bibfnamefont{T.}
  \bibnamefont{Ito}},
  \bibinfo{journal}{Phys. Rev. Lett.}
  \textbf{\bibinfo{volume}{76}}, \bibinfo{pages}{3412} (\bibinfo{year}{1996}).

\bibitem[{\citenamefont{Kivelson et~al.}(2003)\citenamefont{Kivelson, Bindloss,
  Fradkin, Oganesyan, Tranquada, Kapitulnik, and Howald}}]{Kivelson:2003}
\bibinfo{author}{\bibfnamefont{S.~A.} \bibnamefont{Kivelson}},
  \bibinfo{author}{\bibfnamefont{I.~P.} \bibnamefont{Bindloss}},
  \bibinfo{author}{\bibfnamefont{E.}~\bibnamefont{Fradkin}},
  \bibinfo{author}{\bibfnamefont{V.}~\bibnamefont{Oganesyan}},
  \bibinfo{author}{\bibfnamefont{J.~M.} \bibnamefont{Tranquada}},
  \bibinfo{author}{\bibfnamefont{A.}~\bibnamefont{Kapitulnik}},
  \bibnamefont{and} \bibinfo{author}{\bibfnamefont{C.}~\bibnamefont{Howald}},
  \bibinfo{journal}{Rev. Mod. Phys.} \textbf{\bibinfo{volume}{75}},
  \bibinfo{eid}{1201} (\bibinfo{year}{2003}).

\bibitem[{\citenamefont{Tranquada}(2005)}]{Tranquada:2005}
\bibinfo{author}{\bibfnamefont{J.~M.} \bibnamefont{Tranquada}},
  \bibinfo{journal}{to appear as a chapter in ``Treatise of High Temperature
  Superconductivity'' edited by J. Robert Schrieffer; e-print:
  cond-mat/0512115}  (\bibinfo{year}{2005}).

\bibitem[{\citenamefont{Wakimoto et~al.}(1999)\citenamefont{Wakimoto, Shirane,
  Endoh, Hirota, Ueki, Yamada, Birgeneau, Kastner, Lee, Gehring
  et~al.}}]{Wakimoto:1999}
\bibinfo{author}{\bibfnamefont{S.}~\bibnamefont{Wakimoto}},
  \bibinfo{author}{\bibfnamefont{G.}~\bibnamefont{Shirane}},
  \bibinfo{author}{\bibfnamefont{Y.}~\bibnamefont{Endoh}},
  \bibinfo{author}{\bibfnamefont{K.}~\bibnamefont{Hirota}},
  \bibinfo{author}{\bibfnamefont{S.}~\bibnamefont{Ueki}},
  \bibinfo{author}{\bibfnamefont{K.}~\bibnamefont{Yamada}},
  \bibinfo{author}{\bibfnamefont{R.~J.} \bibnamefont{Birgeneau}},
  \bibinfo{author}{\bibfnamefont{M.~A.} \bibnamefont{Kastner}},
  \bibinfo{author}{\bibfnamefont{Y.~S.} \bibnamefont{Lee}},
  \bibinfo{author}{\bibfnamefont{P.~M.} \bibnamefont{Gehring}},
  \bibinfo{journal}{Phys. Rev. B}
  \textbf{\bibinfo{volume}{60}}, \bibinfo{pages}{R769} (\bibinfo{year}{1999}).

  \bibitem[{\citenamefont{Tassini et~al.}(2005)\citenamefont{Tassini, Venturini,
  Zhang, Hackl, Kikugawa, and Fujita}}]{Tassini:2005}
\bibinfo{author}{\bibfnamefont{L.}~\bibnamefont{Tassini}},
  \bibinfo{author}{\bibfnamefont{F.}~\bibnamefont{Venturini}},
  \bibinfo{author}{\bibfnamefont{Q.-M.} \bibnamefont{Zhang}},
  \bibinfo{author}{\bibfnamefont{R.}~\bibnamefont{Hackl}},
  \bibinfo{author}{\bibfnamefont{N.}~\bibnamefont{Kikugawa}}, \bibnamefont{and}
  \bibinfo{author}{\bibfnamefont{T.}~\bibnamefont{Fujita}},
  \bibinfo{journal}{Phys. Rev. Lett.} \textbf{\bibinfo{volume}{95}},
  \bibinfo{eid}{117002} (\bibinfo{year}{2005}).


\bibitem[{\citenamefont{Hoffman et~al.}(2002)\citenamefont{Hoffman, Hudson,
  Lang, Madhavan, Eisaki, Uchida, and Davis}}]{Hoffman:2002}
\bibinfo{author}{\bibfnamefont{J.~E.} \bibnamefont{Hoffman}},
  \bibinfo{author}{\bibfnamefont{K.~M.} \bibnamefont{Lang}},
  \bibinfo{author}{\bibfnamefont{V.} \bibnamefont{Madhaven}},
  \bibinfo{author}{\bibfnamefont{H.}~\bibnamefont{Eisaki}},
  \bibinfo{author}{\bibfnamefont{S.}~\bibnamefont{Uchida}}, , \bibnamefont{and}
  \bibinfo{author}{\bibfnamefont{J.~C.} \bibnamefont{Davis}},
  \bibinfo{journal}{Science} \textbf{\bibinfo{volume}{295}},
  \bibinfo{pages}{466} (\bibinfo{year}{2002}).


\bibitem[{\citenamefont{Hinkov et~al.}(2004)\citenamefont{Hinkov, Pailhes,
  Bourges, Sidis, Ivanov, Kulakov, Lin, Chen, Bernhard, and
  Keimer}}]{Hinkov:2004}
\bibinfo{author}{\bibfnamefont{V.}~\bibnamefont{Hinkov}},
 \bibinfo{author}{\bibfnamefont{S.}~\bibnamefont{Pailhes}},
  \bibinfo{author}{\bibfnamefont{P.}~\bibnamefont{Bourges}},
  \bibinfo{author}{\bibfnamefont{Y.}~\bibnamefont{Sidis}},
  \bibinfo{author}{\bibfnamefont{A.}~\bibnamefont{Ivanov}},
  \bibinfo{author}{\bibfnamefont{A.}~\bibnamefont{Kulakov}},
  \bibinfo{author}{\bibfnamefont{C.~T.} \bibnamefont{Lin}},
  \bibinfo{author}{\bibfnamefont{D.~P.} \bibnamefont{Chen}},
  \bibinfo{author}{\bibfnamefont{C.}~\bibnamefont{Bernhard}}, \bibnamefont{and}
  \bibinfo{author}{\bibfnamefont{B.}~\bibnamefont{Keimer}},
  \bibinfo{journal}{Nature} \textbf{\bibinfo{volume}{430}},
  \bibinfo{pages}{650} (\bibinfo{year}{2004}), ISSN \bibinfo{issn}{0028-0836}.

\bibitem[{\citenamefont{Stock et~al.}(2005)\citenamefont{Stock, Buyers, Cowley,
  Clegg, Coldea, Frost, Liang, Peets, Bonn, Hardy et~al.}}]{Stock:2005}
\bibinfo{author}{\bibfnamefont{C.}~\bibnamefont{Stock}},
  \bibinfo{author}{\bibfnamefont{W.~J.~L.} \bibnamefont{Buyers}},
  \bibinfo{author}{\bibfnamefont{R.~A.} \bibnamefont{Cowley}},
  \bibinfo{author}{\bibfnamefont{P.~S.} \bibnamefont{Clegg}},
  \bibinfo{author}{\bibfnamefont{R.}~\bibnamefont{Coldea}},
  \bibinfo{author}{\bibfnamefont{C.~D.} \bibnamefont{Frost}},
  \bibinfo{author}{\bibfnamefont{R.}~\bibnamefont{Liang}},
  \bibinfo{author}{\bibfnamefont{D.}~\bibnamefont{Peets}},
  \bibinfo{author}{\bibfnamefont{D.}~\bibnamefont{Bonn}},
  \bibinfo{author}{\bibfnamefont{W.~N.} \bibnamefont{Hardy}},
  \bibinfo{journal}{Phys. Rev. B}
  \textbf{\bibinfo{volume}{71}}, \bibinfo{eid}{024522} (\bibinfo{year}{2005}).


\bibitem[{\citenamefont{Hinkov et~al.}(2008)\citenamefont{Hinkov, Haug,
  Fauqu\`e, Bourges, Sidis, Ivanov, Bernhard, Lin, and Keimer}}]{Hinkov:2007}
\bibinfo{author}{\bibfnamefont{V.}~\bibnamefont{Hinkov}},
  \bibinfo{author}{\bibfnamefont{D.}~\bibnamefont{Haug}},
  \bibinfo{author}{\bibfnamefont{B.}~\bibnamefont{Fauqu\`e}},
  \bibinfo{author}{\bibfnamefont{P.}~\bibnamefont{Bourges}},
  \bibinfo{author}{\bibfnamefont{Y.}~\bibnamefont{Sidis}},
  \bibinfo{author}{\bibfnamefont{A.}~\bibnamefont{Ivanov}},
  \bibinfo{author}{\bibfnamefont{C.} \bibnamefont{Bernhard}},
  \bibinfo{author}{\bibfnamefont{C.~T.} \bibnamefont{Lin}},
\bibnamefont{and} \bibinfo{author}{\bibfnamefont{B.}~\bibnamefont{Keimer}},
  \bibinfo{journal}{Science} \textbf{\bibinfo{volume}{319}},
  \bibinfo{pages}{597} (\bibinfo{year}{2008}).

\bibitem[{\citenamefont{Castellani et~al.}(1998)\citenamefont{Castellani,
  Di~Castro, and Grilli}}]{Castellani:1998}
\bibinfo{author}{\bibfnamefont{C.}~\bibnamefont{Castellani}},
  \bibinfo{author}{\bibfnamefont{C.}~\bibnamefont{Di~Castro}},
  \bibnamefont{and} \bibinfo{author}{\bibfnamefont{M.}~\bibnamefont{Grilli}},
  \bibinfo{journal}{J. Phys. Chem. Solids} \textbf{\bibinfo{volume}{59}},
  \bibinfo{pages}{1694} (\bibinfo{year}{1998}).


\bibitem[{\citenamefont{Klauss et~al.}(2000)\citenamefont{Klauss, Wagener,
  Hillberg, Kopmann, Walf, Litterst, H\"ucker, and B\"uchner}}]{Klauss:2000}
\bibinfo{author}{\bibfnamefont{H.-H.} \bibnamefont{Klauss}},
  \bibinfo{author}{\bibfnamefont{W.}~\bibnamefont{Wagener}},
  \bibinfo{author}{\bibfnamefont{M.}~\bibnamefont{Hillberg}},
  \bibinfo{author}{\bibfnamefont{W.}~\bibnamefont{Kopmann}},
  \bibinfo{author}{\bibfnamefont{H.}~\bibnamefont{Walf}},
  \bibinfo{author}{\bibfnamefont{F.~J.} \bibnamefont{Litterst}},
  \bibinfo{author}{\bibfnamefont{M.}~\bibnamefont{H\"ucker}}, \bibnamefont{and}
  \bibinfo{author}{\bibfnamefont{B.}~\bibnamefont{B\"uchner}},
  \bibinfo{journal}{Phys. Rev. Lett.} \textbf{\bibinfo{volume}{85}},
  \bibinfo{pages}{4590} (\bibinfo{year}{2000}).


\bibitem[{\citenamefont{Devereaux and Hackl}(2007)}]{Devereaux:2007}
\bibinfo{author}{\bibfnamefont{T.~P.} \bibnamefont{Devereaux}}
  \bibnamefont{and} \bibinfo{author}{\bibfnamefont{R.}~\bibnamefont{Hackl}},
  \bibinfo{journal}{Rev. Mod. Phys.} \textbf{\bibinfo{volume}{79}},
  \bibinfo{pages}{175} (\bibinfo{year}{2007}).

\bibitem[{\citenamefont{Opel et~al.}(2000)\citenamefont{Opel, Nemetschek,
  Hoffmann, Philipp, M\"uller, Hackl, T\"utt\H{o}, Erb, Revaz, Walker
  et~al.}}]{Opel:2000}
\bibinfo{author}{\bibfnamefont{M.}~\bibnamefont{Opel}},
  \bibinfo{author}{\bibfnamefont{R.}~\bibnamefont{Nemetschek}},
  \bibinfo{author}{\bibfnamefont{C.}~\bibnamefont{Hoffmann}},
  \bibinfo{author}{\bibfnamefont{R.}~\bibnamefont{Philipp}},
  \bibinfo{author}{\bibfnamefont{P.~F.} \bibnamefont{M\"uller}},
  \bibinfo{author}{\bibfnamefont{R.}~\bibnamefont{Hackl}},
  \bibinfo{author}{\bibfnamefont{I.}~\bibnamefont{T\"utt\H{o}}},
  \bibinfo{author}{\bibfnamefont{A.}~\bibnamefont{Erb}},
  \bibinfo{author}{\bibfnamefont{B.}~\bibnamefont{Revaz}},
  \bibinfo{author}{\bibfnamefont{E.}~\bibnamefont{Walker}},
  \bibnamefont{et~al.},
  \bibinfo{journal}{Phys. Rev. B}
  \textbf{\bibinfo{volume}{61}}, \bibinfo{pages}{9752} (\bibinfo{year}{2000}).

  \bibitem[{\citenamefont{Caprara et~al.}(2005)\citenamefont{Caprara, Di~Castro,
  Grilli, and Suppa}}]{Caprara:2005}
\bibinfo{author}{\bibfnamefont{S.}~\bibnamefont{Caprara}},
  \bibinfo{author}{\bibfnamefont{C.}~\bibnamefont{Di Castro}},
  \bibinfo{author}{\bibfnamefont{M.}~\bibnamefont{Grilli}}, \bibnamefont{and}
  \bibinfo{author}{\bibfnamefont{D.}~\bibnamefont{Suppa}},
  \bibinfo{journal}{Phys. Rev. Lett.} \textbf{\bibinfo{volume}{95}},
  \bibinfo{eid}{117004} (\bibinfo{year}{2005}).

\bibitem[{\citenamefont{Erb et~al.}(1996)\citenamefont{Erb, Walker, and
  Fl\"ukiger}}]{Erb:1996}
\bibinfo{author}{\bibfnamefont{A.}~\bibnamefont{Erb}},
  \bibinfo{author}{\bibfnamefont{E.}~\bibnamefont{Walker}}, \bibnamefont{and}
  \bibinfo{author}{\bibfnamefont{R.}~\bibnamefont{Fl\"ukiger}},
  \bibinfo{journal}{Physica C} \textbf{\bibinfo{volume}{258}},
  \bibinfo{pages}{9} (\bibinfo{year}{1996}).

\bibitem[{\citenamefont{J\'anossy et~al.}(2003)\citenamefont{J\'anossy,
  Feh\'er, and Erb}}]{Janossy:2003}
\bibinfo{author}{\bibfnamefont{A.}~\bibnamefont{J\'anossy}},
  \bibinfo{author}{\bibfnamefont{T.}~\bibnamefont{Feh\'er}}, \bibnamefont{and}
  \bibinfo{author}{\bibfnamefont{A.}~\bibnamefont{Erb}},
  \bibinfo{journal}{Phys. Rev. Lett.} \textbf{\bibinfo{volume}{91}},
  \bibinfo{eid}{177001} (\bibinfo{year}{2003}).


\bibitem[{\citenamefont{Sugai et~al.}(2003)\citenamefont{Sugai, Suzuki,
  Takayanagi, Hosokawa, and Hayamizu}}]{Sugai:2003}
\bibinfo{author}{\bibfnamefont{S.}~\bibnamefont{Sugai}},
  \bibinfo{author}{\bibfnamefont{H.}~\bibnamefont{Suzuki}},
  \bibinfo{author}{\bibfnamefont{Y.}~\bibnamefont{Takayanagi}},
  \bibinfo{author}{\bibfnamefont{T.}~\bibnamefont{Hosokawa}}, \bibnamefont{and}
  \bibinfo{author}{\bibfnamefont{N.}~\bibnamefont{Hayamizu}},
  \bibinfo{journal}{Phys. Rev. B} \textbf{\bibinfo{volume}{68}},
  \bibinfo{pages}{184504} (\bibinfo{year}{2003}).

\bibitem[{\citenamefont{Gozar et~al.}(2005)\citenamefont{Gozar, Komiya, Ando,
  and Blumberg}}]{Gozar:2005}
\bibinfo{author}{\bibfnamefont{A.}~\bibnamefont{Gozar}},
  \bibinfo{author}{\bibfnamefont{S.}~\bibnamefont{Komiya}},
  \bibinfo{author}{\bibfnamefont{Y.}~\bibnamefont{Ando}}, \bibnamefont{and}
  \bibinfo{author}{\bibfnamefont{G.}~\bibnamefont{Blumberg}},
  \emph{\bibinfo{title}{Frontiers in Magnetic Materials}}
  (\bibinfo{publisher}{Springer-Verlag, Berlin, p. 755}, \bibinfo{year}{2005}).

\bibitem[{\citenamefont{Ando et~al.}(2004)\citenamefont{Ando, Komiya, Segawa,
  Ono, and Kurita}}]{Ando:2004}
\bibinfo{author}{\bibfnamefont{Y.}~\bibnamefont{Ando}},
  \bibinfo{author}{\bibfnamefont{S.}~\bibnamefont{Komiya}},
  \bibinfo{author}{\bibfnamefont{K.}~\bibnamefont{Segawa}},
  \bibinfo{author}{\bibfnamefont{S.}~\bibnamefont{Ono}}, \bibnamefont{and}
  \bibinfo{author}{\bibfnamefont{Y.}~\bibnamefont{Kurita}},
  \bibinfo{journal}{Phys. Rev. Lett.} \textbf{\bibinfo{volume}{93}},
  \bibinfo{eid}{267001} (\bibinfo{year}{2004}).


\bibitem[{\citenamefont{Venturini et~al.}(2002)\citenamefont{Venturini, Opel,
  Devereaux, Freericks, T\"utt\H{o}, Revaz, Walker, Berger, Forr\'o, and
  Hackl}}]{Venturini:2002}
\bibinfo{author}{\bibfnamefont{F.}~\bibnamefont{Venturini}},
  \bibinfo{author}{\bibfnamefont{M.}~\bibnamefont{Opel}},
  \bibinfo{author}{\bibfnamefont{T.~P.} \bibnamefont{Devereaux}},
  \bibinfo{author}{\bibfnamefont{J.~K.} \bibnamefont{Freericks}},
  \bibinfo{author}{\bibfnamefont{I.}~\bibnamefont{T\"utt\H{o}}},
  \bibinfo{author}{\bibfnamefont{B.}~\bibnamefont{Revaz}},
  \bibinfo{author}{\bibfnamefont{E.}~\bibnamefont{Walker}},
  \bibinfo{author}{\bibfnamefont{H.}~\bibnamefont{Berger}},
  \bibinfo{author}{\bibfnamefont{L.}~\bibnamefont{Forr\'o}}, \bibnamefont{and}
  \bibinfo{author}{\bibfnamefont{R.}~\bibnamefont{Hackl}},
  \bibinfo{journal}{Phys. Rev. Lett.} \textbf{\bibinfo{volume}{89}},
  \bibinfo{pages}{107003} (\bibinfo{year}{2002}).

\bibitem[{\citenamefont{Stock et~al.}(2006)\citenamefont{Stock, Buyers, Yamani,
  Broholm, Chung, Tun, Liang, Bonn, Hardy, and Birgeneau}}]{Stock:2006}
\bibinfo{author}{\bibfnamefont{C.}~\bibnamefont{Stock}},
  \bibinfo{author}{\bibfnamefont{W.~J.~L.} \bibnamefont{Buyers}},
  \bibinfo{author}{\bibfnamefont{Z.}~\bibnamefont{Yamani}},
  \bibinfo{author}{\bibfnamefont{C.~L.} \bibnamefont{Broholm}},
  \bibinfo{author}{\bibfnamefont{J.-H.} \bibnamefont{Chung}},
  \bibinfo{author}{\bibfnamefont{Z.}~\bibnamefont{Tun}},
  \bibinfo{author}{\bibfnamefont{R.}~\bibnamefont{Liang}},
  \bibinfo{author}{\bibfnamefont{D.}~\bibnamefont{Bonn}},
  \bibinfo{author}{\bibfnamefont{W.~N.} \bibnamefont{Hardy}}, \bibnamefont{and}
  \bibinfo{author}{\bibfnamefont{R.~J.} \bibnamefont{Birgeneau}},
  \bibinfo{journal}{Phys. Rev. B} \textbf{\bibinfo{volume}{73}},
  \bibinfo{eid}{100504 (R)} (\bibinfo{year}{2006}).

\bibitem{Tassini:2008}
  L. Tassini, Doctoral Thesis, Technical University Munich (2008).

\bibitem[{\citenamefont{Sushkov and Kotov}(2005)}]{Sushkov:2005}
\bibinfo{author}{\bibfnamefont{O.~P.} \bibnamefont{Sushkov}} \bibnamefont{and}
  \bibinfo{author}{\bibfnamefont{V.~N.} \bibnamefont{Kotov}},
  \bibinfo{journal}{Phys. Rev. Lett.} \textbf{\bibinfo{volume}{94}},
  \bibinfo{eid}{097005} (\bibinfo{year}{2005}).

\bibitem[{\citenamefont{Gr\"uner}(1994)}]{Gruner:1994}
\bibinfo{author}{\bibfnamefont{G.}~\bibnamefont{Gr\"uner}},
  \emph{\bibinfo{title}{Density Waves in Solids}}
  (\bibinfo{publisher}{Addison-Wesley, Reading, MA}, \bibinfo{year}{1994}).

\bibitem[{\citenamefont{Reznik et~al.}(2006)\citenamefont{Reznik, Pintschovius,
  Ito, Iikubo, Sato, Goka, Fujita, Yamada, Gu, and Tranquada}}]{Reznik:2006}
\bibinfo{author}{\bibfnamefont{D.}~\bibnamefont{Reznik}},
  \bibinfo{author}{\bibfnamefont{L.}~\bibnamefont{Pintschovius}},
  \bibinfo{author}{\bibfnamefont{M.}~\bibnamefont{Ito}},
  \bibinfo{author}{\bibfnamefont{S.}~\bibnamefont{Iikubo}},
  \bibinfo{author}{\bibfnamefont{M.}~\bibnamefont{Sato}},
  \bibinfo{author}{\bibfnamefont{H.}~\bibnamefont{Goka}},
  \bibinfo{author}{\bibfnamefont{M.}~\bibnamefont{Fujita}},
  \bibinfo{author}{\bibfnamefont{K.}~\bibnamefont{Yamada}},
  \bibinfo{author}{\bibfnamefont{G.~D.} \bibnamefont{Gu}}, \bibnamefont{and}
  \bibinfo{author}{\bibfnamefont{J.~M.} \bibnamefont{Tranquada}},
  \bibinfo{journal}{Nature} \textbf{\bibinfo{volume}{440}},
  \bibinfo{pages}{1170} (\bibinfo{year}{2006}).

\bibitem[{\citenamefont{Perali et~al.}(2000)\citenamefont{Perali, Castellani,
  Di~Castro, Grilli, Piegari, and Varlamov}}]{Perali:2000}
\bibinfo{author}{\bibfnamefont{A.}~\bibnamefont{Perali}},
 \bibinfo{author}{\bibfnamefont{C.}~\bibnamefont{Castellani}},
  \bibinfo{author}{\bibfnamefont{C.}~\bibnamefont{Di~Castro}},
  \bibinfo{author}{\bibfnamefont{M.}~\bibnamefont{Grilli}},
  \bibinfo{author}{\bibfnamefont{E.}~\bibnamefont{Piegari}}, \bibnamefont{and}
  \bibinfo{author}{\bibfnamefont{A.~A.} \bibnamefont{Varlamov}},
  \bibinfo{journal}{Phys. Rev. B} \textbf{\bibinfo{volume}{62}},
  \bibinfo{pages}{R9295} (\bibinfo{year}{2000}).

\bibitem[{\citenamefont{Metzner et~al.}(2003)\citenamefont{Metzner, Rohe, and
  Andergassen}}]{Metzner:2003}
\bibinfo{author}{\bibfnamefont{W.}~\bibnamefont{Metzner}},
  \bibinfo{author}{\bibfnamefont{D.}~\bibnamefont{Rohe}}, \bibnamefont{and}
  \bibinfo{author}{\bibfnamefont{S.}~\bibnamefont{Andergassen}},
  \bibinfo{journal}{Phys. Rev. Lett.} \textbf{\bibinfo{volume}{91}},
  \bibinfo{eid}{066402} (\bibinfo{year}{2003}).

\bibitem[{\citenamefont{Seibold et~al.}(2007)\citenamefont{Seibold, Lorenzana,
  and Grilli}}]{Seibold:2006}
\bibinfo{author}{\bibfnamefont{G.}~\bibnamefont{Seibold}},
  \bibinfo{author}{\bibfnamefont{J.}~\bibnamefont{Lorenzana}},
  \bibnamefont{and} \bibinfo{author}{\bibfnamefont{M.}~\bibnamefont{Grilli}},
  \bibinfo{journal}{Phys. Rev. B} \textbf{\bibinfo{volume}{75}}, \bibinfo{eid}{100505 (R)},
   (\bibinfo{year}{2007}).

\end{thebibliography}

\end{document}